\documentstyle[12pt,a4]{article}

\newcommand{\bea}{\begin{eqnarray}}
\newcommand{\eea}{\end{eqnarray}}
\newcommand{\be}{\begin{equation}}
\newcommand{\ee}{\end{equation}}

\begin{document}

\begin{titlepage}

\begin{flushright}
FTUV-99-61, IFIC/99-64
\end{flushright}

\baselineskip 24pt

\begin{center}

{\Large {\bf Coherent Muon-Electron Conversion in the Dualized 
Standard Model }}\\

\vspace{.5cm}

\baselineskip 14pt
{\large Jos\'e BORDES}\\
jose.m.bordes\,@\,uv.es\\
{\it Departament Fisica Teorica, Universitat de Valencia,\\
  calle Dr. Moliner 50, E-46100 Burjassot (Valencia), Spain}\\
\vspace{.2cm}
{\large CHAN Hong-Mo}\\
chanhm\,@\,v2.rl.ac.uk  \,\,\,  \\
{\it Rutherford Appleton Laboratory,\\
  Chilton, Didcot, Oxon, OX11 0QX, United Kingdom}\\
\vspace{.2cm}
{\large Ricardo GALLEGO}\\
{\it Departament Fisica Teorica, Universitat de Valencia,\\
  calle Dr. Moliner 50, E-46100 Burjassot (Valencia), Spain}\\
\vspace{.2cm}
{\large TSOU Sheung Tsun}\\
tsou\,@\,maths.ox.ac.uk\\
{\it Mathematical Institute, University of Oxford,\\
  24-29 St. Giles', Oxford, OX1 3LB, United Kingdom}
\end{center}

\vspace{.3cm}

\begin{abstract}
Muon-electron conversion in nuclei is considered in the framework of the 
Dualized Standard Model. The ratio $B_{\mu-e}$ of the conversion rate to 
the total muon capture rate is derived, and computed for several nuclei 
in a parameter-free calculation using parameters previously determined 
in different physical contexts.  The values obtained all lie within the 
present experimental bounds, but some are so close as to seem readily 
accessible to experiments already being planned.  Similar considerations 
are applied also to muon-electron conversion in muonium but give rates 
many orders of magnitude below the present experiment limit. 
\end{abstract}

\end{titlepage}

\clearpage

In this note we present some results of the Dualized Standard Model (DSM) for 
coherent muon-electron conversion in the nuclear elastic (ground state to
ground state) process:
\be
\mu^- + (Z,N) \rightarrow e^- + (Z,N).
\label{nuclearconv}
\ee

This process entails a change of flavor number and is forbidden in the
conventional version of the Standard Model.  It is allowed however in many 
extensions of Standard Model considered in the literature with so-called 
``horizontal'' symmetries linking the different generations.  (For a 
general discussion of the problem see \cite{cahnharari}).   For this 
reason, it is widely accepted that precision experiments on muon-electron 
conversion are an excellent framework to test theories beyond the 
conventional Standard Model.  

Present experiments give the best limits on the ratio of the conversion 
rate to the muon capture rate as \cite{databook}:
\bea
B^{Pb}_{\mu-e} & \leq  &  4.6 \,\times \, 10^{-11} 
\nonumber \\
B^{Ti}_{\mu-e} & \leq & 4.3 \,\times\, 10^{-12}
\label{experimental}
\eea
which put stringent limits on flavor-changing currents.  Furthermore, 
experimental sensitivity is expected to be improved by three or four 
orders of magnitude by the planned experiment at PSI \cite{psi} using 
a $^{48}Ti$ target, and the MECO experiment \cite{meco} at Brookhaven, 
using a $^{27}Al$ target.  They will thus afford very valuable tests for 
extensions beyond the conventional Standard Model.

However, most extensions of the Standard Model (SUSY, GUT's...) involve 
ingredients outside the theoretical framework of the Standard Model.  
Their predictions on muon-conversion processes would thus generally depend 
on various ``external'' parameters (symmetry breaking scales, mixing 
matrices...) which are not given by the theory but have to be determined 
from experiment on non-standard effects, so that their predictive power 
is in general limited.  In contrast, the DSM scheme suggested in Ref. 
\cite{chantsou2} relies on an earlier result that within the Standard 
Model itself there is already a dual color group which can play the role 
of the horizontal symmetry of generations.  Most of the parameters in 
the scheme are thus either already given by the theory or else determined 
by fitting standard quantities such as masses and mixing angles, so 
that actual predictions for non-standard processes can now be given in 
a quite unambiguous manner. 

Our aim in this paper is to examine the predictions of the DSM scheme on
muon-conversion processes using the formalism and techniques developed
previously for an analysis on flavor-changing neutral currents effects in
meson mass differences and rare meson decays \cite{FCNC}.  Muon conversion 
rates are then obtained, which turn out to be quite close to the present 
experimental limits and should thus be testable by experiment in the 
very near future.

We begin with a brief outline of the basic tenets of the DSM scheme.  Based 
on a non-Abelian generalization of electric-magnetic duality \cite{chantsou1}
and a result on confinement of 't Hooft's \cite{thooft}, the dualized version 
of the Standard Model has been constructed \cite{chantsou2} which offers 
an explanation for the existence of just three generations of fermions as 
a broken local dual color gauge symmetry, and also of Higgs fields as frame 
vectors in dual color space.  This dual color $\widetilde{SU}(3)$ group 
which is identified with the generation symmetry, is spontaneously broken in 
such a way that, at the one-loop level of dual boson-exchange, the (real) 
mixing matrices and masses can be calculated for both quarks and leptons.  The
results obtained are in good agreement with present experiment.  A full 
account of the whole process is given in \cite{ckm,nuos,phenodsm,interim}
where it was shown that with only three free parameters (the two ratios between
the three dual Higgs v.e.v.'s plus a common Yukawa coupling) one is able 
to reproduce to an acceptable level some 14 of the ``fundamental'' parameters 
of the Standard Model.

We note in particular the following features of the DSM scheme which are 
of special relevance to the problem at hand.  First, by virtue of the Dirac 
quantization condition, no unknown gauge couplings appear in the DSM.  The 
gauge couplings $\tilde{g}_i$ of the dual groups are related to those of 
the ``direct'' groups $g_i$ \cite{chantsou3}, namely the ordinary color and 
electroweak gauge couplings routinely measured in present day experiments:
$$
g_{3(2)} \tilde{g}_{3(2)} = 4\pi, \;\;\; g_1 \tilde{g}_1 = 2\pi.
$$
In other words, the coupling strengths $\tilde{g}_3$ and $\tilde{g}_1$ of 
the dual gauge bosons, which we shall need in what follows for considerations 
of muon conversion, are derivable from respectively the usual strong 
coupling constant $g_3$ and the coupling of weak hypercharge $g_1$.

Secondly, the branching of these couplings $\tilde{g}_i$ into the various 
physical fermion states are given by the orientations of these physical 
states in generation or dual color space.  If we denote by $\psi_{L}^A$ 
the left-handed fermion fields of type $A$, the relation between the 
$physical$ and $gauge$ basis in generation space is given by a unitary 
matrix $S^A$:
\begin{equation}
\psi_{gauge,L}^A = S^A \psi_{physical,L}^A
\label{rotationf}
\end{equation} 
where the index $A$ runs over the four types of fermions $U, D, L$ (charged 
leptons) and $N$ (neutrinos).  These orientation matrices $S^A$ were already 
determined as by-products in the calculation of fermion mixing matrices 
\cite{ckm,nuos} by fitting the free parameters of the model to the masses 
of the higher generation fermions and to the Cabibbo angle.  Those matrices 
for quarks and leptons which are relevant for later development in this paper 
are given in \cite{FCNC}.  We note that in terms of $S^A$, the CKM matrix 
for leptons (quarks) is given simply as $V^{lepton}_{CKM}=(S^N)^\dagger S^L$ 
($V^{quark}_{CKM}=(S^U)^\dagger S^D$), and the result on the mixing matrices 
of \cite{ckm,nuos,phenodsm} was found to be in excellent agreement with 
present experimental values. 

Thirdly, the fit done in \cite{ckm} quoted above also gave a hierarchical 
relation between the v.e.v.'s of the Higgs bosons responsible for
breaking the dual color or generation symmetry, which implies in turn 
that in the tree-level spectrum found for the dual color gauge bosons, 
one particular state has a much lower mass than the rest, so that the 
calculation of the low energy effective Lagrangian relevant for one-dual 
gauge boson exchange becomes quite simple, being dominated by just the 
exchange of this one boson.  We give here the expression for the full 
effective Lagrangian \cite{FCNC}, 
\begin{equation}
L_{eff}=
\frac{1}{2 (\zeta z)^2} 
\sum_{A,B} f^{A,B}_{\alpha,\beta;\alpha',\beta'} 
(J^{\mu \dagger}_A)^{\alpha,\beta}  (J_{\mu,B})^{\alpha',\beta'}\,,
\label{laeff1}
\end{equation}
with currents of the usual $\;V-A\;$ form:
$$
(J^{A}_\mu)_{\alpha,\beta} = \bar{\psi}_{L,\alpha}^A \gamma_\mu 
   \psi_{L,\beta}^A,
$$
where Greek indices run over flavors, and the group factor, in the present
case in which only the lightest gauge boson is relevant, reduces to the
following particularly simple combination of the rotating matrices
(\ref{rotationf}):
\begin{equation}
f^{A,B}_{\alpha,\beta;\alpha',\beta'} =
S^{A*}_{3,\alpha} S^{A}_{3,\beta}
S^{B*}_{3,\alpha'} S^{B}_{3,\beta'} 
\label{groupfactor1}
\end{equation}
The structure of this $L_{eff}$ is self explanatory in the sense that it 
links together fermions of equal or different flavors (A and B).

Finally, the only remaining unknown among the quantities required which
appeared in (\ref{laeff1}) is the mass scale $\zeta z$ of the breaking of 
dual color symmetry. This parameter was not constrained by the calculation
of mixing matrices in \cite{ckm,nuos,phenodsm} but are bounded by other
considerations \cite{FCNC,airshower1} as follows.  By studying the effects
of dual color symmetry breaking on FCNC meson decay and on the mass 
differences of conjugate neutral meson pairs, one finds that the mass 
difference $K_L - K_S$ is the most restrictive, giving a lower bound 
for the scale $\zeta z$ of the order of a few hundred Tev \cite{FCNC}.
On the other hand, accepting the suggestion \cite{airshower1} that those
rare air shower events with energies beyond the GZK cut-off \cite{greitsemin} 
are produced by neutrinos acquiring strong interactions through the exchange
of dual gauge bosons, one can give a rough upper bound to the scale $\zeta z$
also of the order of a few hundred TeV \cite{FCNC}.  All in all, our best 
guess for this scale $\zeta z$, which in the DSM scheme is the mass of the
lightest dual Higgs boson, is in the range $300-500$ TeV.  This is 
much lower than the scales appearing in other models beyond the SM such 
as GUT's.  Notice, however, that these quoted limits for $\zeta z$ are 
merely rough estimates, especially the upper bound which was deduced 
from the neutrino explanation of post-GZK air showers by only qualitative 
arguments and awaits a detailed analysis when more data become available.
Nevertheless, these estimates will give one some indications for the sort of
values to be expected.

With these tools in hand and all parameters fixed, we now turn to the problem
of muon-electron conversion in nuclei.  We shall first extract from 
(\ref{laeff1}) the piece relevant for these transitions. The result will be 
just the product of two currents, firstly a leptonic piece made out of single 
left-handed lepton fields with different flavors to describe the muon-electron 
conversion, and secondly a flavor-conserving hadronic current corresponding
to the fact that the initial and final nuclei have the same number of 
neutrons and protons.  This is multiplied by an effective coupling strength 
which is essentially the inverse square of the scale $\zeta z$ times a group 
factor $f^{A,B}_{\alpha,\beta;\alpha',\beta'}$ coming from the rotation 
between gauge and physical states in generation or dual color space 
(\ref{groupfactor1}).  Explicitly, at the quark and lepton level, the 
relevant piece of the effective Lagrangian for the conversion processes 
of interest in this paper is as follows\footnote{Note that the ordering 
convention of the generation indices here is from high to low mass.}:
\begin{equation}
L^{L,Q}_{eff} = \frac{1}{ (\zeta z)^2 } 
\left(  e_L \gamma^\mu  \bar{\mu}_L \right)
\left\{ 
f^{L,U}_{3,2;3,3} \bar{u}_L \gamma_\mu u_L +
f^{L,D}_{3,2;3,3} \bar{d}_L \gamma_\mu d_{L} 
\right\},
\label{leffconversion}
\end{equation}
This embodies all the elementary information we need for computing the 
conversion probability in the one dual gauge boson exchange approximation.

The next step is to find the effective Lagrangian appropriate for the nuclear
process (\ref{nuclearconv}), i.e. to pass from the Lagrangian given in terms 
of the elementary quark fields to a Lagrangian in terms of neutrons and 
protons.  Usually, the passage from quarks to nucleons is effected by using 
the Lorentz invariant form factors depending on the momentum transfer $(q^2)$,
which provide a phenomenological parametrization of the non-perturbative 
structure of the nucleons as quark bound states, and can be largely
determined through experimental observations and symmetry considerations. 
In the present case, the momentum transfer is very small compared to any 
other mass scale involved in the process, so that one can ignore the terms 
which are suppressed by factors of $(q^2)$ over the nucleus mass, as well 
as the dependence on $(q^2)$.  Hence one can just keep the contributions 
of the vector and axial vector form factors $G_{V,A}(q^2)$ taken at $q^2=0$, 
which leads to the following approximation for the nuclear currents:
\bea
\langle \phi| \bar{q} \gamma_\mu q |\phi \rangle & = & G_{V}^{q,N} 
   \bar{\phi} \gamma_\mu \phi,  \,\,\,  \nonumber \\
\langle \phi| \bar{q} \gamma_\mu \gamma_5 q |\phi \rangle & = & G_{A}^{q,N} 
   \bar{\phi} \gamma_\mu \gamma_5 \phi,  \,\,\,  \nonumber
\eea
where $\phi$ represents the wave function of a single nucleon $N$, which
can be either a proton $p$ or a neutron $n$.

In the limit of exact isospin symmetry, the vector piece of the nucleon 
form factors is given by:
\bea
& & G_{V}^{u,n}=G_{V}^{d,p}=G_{V}^{d}=1,
\nonumber \\
& & G_{V}^{u,p}=G_{V}^{d,n}=G_{V}^{u}=2.
\nonumber
\eea
The last equality in both cases comes from considering a coherent contribution
to the scattering process from the valence quarks inside the nucleon.  The
same argument cannot be directly applied to the axial-vector piece but, for 
practical reasons this is irrelevant to us: being dependent on the nuclear 
spin (in the non-relativistic limit) it is incoherent from the point of view
of the nucleons.  Only nucleons outside closed shells contribute to this piece
and it is much smaller than the vectorial, coherent, contribution\footnote{For
a detailed discussion of the axial-vector piece, see e.g. \cite{bernabeu}.}.  
Thus, for our purpose we shall just parametrize this contribution in a 
generic way by the axial vector form factor $g_A^{(N)}$.  The effective 
Lagrangian in terms of proton and neutron fields then reads as:
\bea
& &
L^{L,Q}_{eff} = \frac{1}{ (\zeta z)^2 } 
\left( e_L \gamma^\mu \bar{\mu}_L \right)
\left\{ 
\left( 2 f^{L,U}_{3,2;3,3} + f^{L,D}_{3,2;3,3} \right)
\left( \bar{p} \gamma_\mu \frac{1-g_A^{(p)} \gamma_5}{2} p \right)
\right.
\nonumber \\
& + &
\left.
\left( f^{L,U}_{3,2;3,3} + 2 f^{L,D}_{3,2;3,3} \right)
\left( \bar{n} \gamma_\mu \frac{1-g_A^{(n)} \gamma_5}{2} n \right)
\right\},
\label{leff5}
\eea
where $p$ and $n$ correspond to the spinor fields for respectively protons 
and neutrons.

It is sometimes useful in nuclear processes to use the isospin formalism and
write the fields in an isospin doublet ($N$).  For completeness, we give the
form of (\ref{leff5}) also in this notation:
\bea
& &
L^{L,N}_{eff} = \frac{1}{ (\zeta z)^2 } 
\left(e_L \gamma^\mu  \bar{\mu}_L \right)
\left\{ 
\frac{3}{2} \left( f^{L,U}_{3,2;3,3} + f^{L,D}_{3,2;3,3} \right)
\left( \bar{N} \gamma_\mu \frac{1-g_A^{(p)'} \gamma_5}{2} N \right)
\right.
\nonumber \\
& + &
\left.
\frac{1}{2} \left( f^{L,U}_{3,2;3,3} -  f^{L,D}_{3,2;3,3} \right)
\left( \bar{N} \gamma_\mu \frac{1-g_A^{(n)'} \gamma_5}{2}  \tau_3 N \right)
\right\},
\nonumber
\eea
where $g_A^{(N)'}$ is an averaged axial form factor including both $g_A^{(p)}$
and $g_A^{(n)}$ \cite{bernabeu}.  Its explicit form is not important here.

Next, using the non-relativistic approximation for the nuclear motion, which
allows us to consider only the large component of the nucleon wave function 
(notice that the small component is $O(\vec{p}/M)$ for a nucleon of mass 
$M$ and momentum $\vec{p}$), we can perform the following substitution in 
the Dirac $\gamma$ matrices:  $\gamma^0 \rightarrow I $, $\vec{\gamma}\sim 
\gamma^0 \gamma^5 \sim \vec{p}/M \rightarrow 0 $, $\vec{\gamma}\gamma^5 
\rightarrow \vec{\sigma}$ when taken between hadron states.  The effective 
Lagrangian can then be expressed in the following form:
\bea
L^{L,N}_{eff} & = & 
\frac{1}{2 (\zeta z)^2 }
\left\{
\left( e^+_L \mu_L \right)
\left[
\frac{3}{2} \left( f^{L,U}_{3,2;3,3} + f^{L,D}_{3,2;3,3} \right)
\left(N^+ N \right) +
\right.  \right. 
\nonumber \\ 
& + &
\left.
\frac{1}{2} \left( f^{L,U}_{3,2;3,3} -  f^{L,D}_{3,2;3,3} \right)
\left( N^+ \tau_3  N \right)
\right] -
\nonumber \\ 
& - &
\left( \bar{e}_L \vec{\gamma} \mu_L \right)
\left[
g_A \frac{3}{2} \left( f^{L,U}_{3,2;3,3} + f^{L,D}_{3,2;3,3} \right)
\left( N^+ \vec{\sigma} N  \right) + 
\right.
\nonumber \\
& + &
\left. \left.
g_A' \frac{1}{2} \left( f^{L,U}_{3,2;3,3} -  f^{L,D}_{3,2;3,3} \right)
\left( N^+ \vec{\sigma} \tau_3  N \right)
\right]
\right\},
\nonumber
\eea

Further, as already mentioned above, the $\vec{\sigma}$ dependent term is 
proportional to the spin of the nucleus which, in turn, depends on the 
number of nucleons outside closed shells.  In middle and heavy nuclei, this 
number is much smaller than the atomic weight $A$.  On the other hand the 
$\tau_3$ terms can be interpreted as the isospin density which is comparable 
to $(Z-N)\rho(x)$, $\rho(x)$ being the average density for protons and 
neutrons inside the nucleus.  Compared to the atomic weight, this term is 
again negligible for middle nuclei where $Z$ and $N$ are comparable.  Hence,
for the nuclei we shall be most interested in, the only remaining term is
$N^+ N$ which in the position configuration is directly proportional to 
the nuclear density.  Normalizing $\rho$ to unity, we can then make the 
substitution:
$$
N^+ N \rightarrow A \rho(x).
$$
As a result, we get the final form of the effective Lagrangian relevant
to the muon-electron conversion process, which simply reads as:
\be
L^{L,N}_{eff} =  \frac{1}{(\zeta z)^2} Q \rho(x) e^+_L \mu_L,
\label{finalLeff}
\ee
where
\be
Q= \frac{3 A}{4} 
\left( f^{L,U}_{3,2;3,3} + f^{L,D}_{3,2;3,3} \right).
\label{charge}
\ee

It is interesting to note in (\ref{finalLeff}) that all the information 
which depends on the flavor-violation scheme, in our case the DSM, is 
contained just in the factor $Q/(\zeta z)^2$.  The remaining part in 
(\ref{finalLeff}) depends only on the (approximated) nuclear dynamics. This 
factor $Q/(\zeta z)^2$ here thus plays the role of a {\it flavor-violating
charge} in giving the flavor-violating coupling of the lepton to the nucleus 
in the one dual gauge boson exchange approximation.  One can thus translate 
the result of (\ref{finalLeff}) directly to any other flavor-changing model 
mediated by one boson exchange just by the substitution of the {\it 
flavor-violating charge} appropriate to that model.
 
In the expression (\ref{charge}) for $Q$, we have assumed that the numbers 
of protons $(Z)$ and neutrons $(N)$ are roughly the same.  More generally,
we can rewrite the expression keeping track of these numbers as follows:
\be
Q= \frac{1}{2} 
\left[ Z \left( 2 f^{L,U}_{3,2;3,3} + f^{L,D}_{3,2;3,3} \right) +
N \left( f^{L,U}_{3,2;3,3} + 2 f^{L,D}_{3,2;3,3} \right) \right],
\label{chargeagain}
\ee
where for simplicity, we have taken similar proton and neutron densities
inside the nucleus.  In general, it will depend weakly on the nuclear 
parameters, essentially through the ratio $\frac{A-2 Z}{A}$.

Substituting next the effective Lagrangian (\ref{finalLeff}) into the 
expression calculated from the scattering process, one obtains for the
conversion rate:
\be
\Gamma_{conv.} = \frac{p_e E_e}{2 \pi}  \frac{1}{(\zeta z)^4} Q^2 |M(q^2)|^2.
\label{muconversionrate}
\ee
where $p_e$ and $E_e$ are the momentum and energy of the electron which, in
the present low energy case, are both of the order of the muon mass.  In the
formula (\ref{muconversionrate}), we have taken the non-relativistic limit 
for the motion of the muon in the muonic atom and in this way we can factorize
the large component of the muon wave function, $\phi_\mu (x)$.  $M(q^2)$ is 
then the Fourier transform of the nucleus density modulated by the muon 
wave function:
\be
M(q^2)= \int d^3x \, \, \rho(x) e^{-i\vec{q}\vec{x}} \phi_\mu (x).
\label{em}
\ee

Now, for nuclei with $A \leq 100$ it is costumary in $\mu$-capture analysis 
to take an average value for the muon wave function inside the nucleus and 
parametrize the nuclear size effects by the form factor in such a way that 
eq. (\ref{em}) becomes
\be
|M(q^2)|^2 = \frac{\alpha^3 m_\mu^3}{\pi}
\frac{Z_{eff}^4}{Z}|F(q^2)|^2,
\label{m2}
\ee
where $Z_{eff}$ has been determined in the literature \cite{shanker,chiang}
and is given by essentially the square of the muon wave function averaged 
over the nucleus, and $F(q^2)$ is the nuclear form factor which can be 
measured for instance in electron scattering.  For our purposes we can 
approximate $F(q^2)$ by a dipole form factor which reproduces the mean 
squared radius of the nucleus, thus:
\be
F(q^2)=\frac{1}{1-\frac{q^2 \langle r^2 \rangle}{6}}
\label{formfactor}
\ee
where the mean squared radius is approximately given by $\langle r^2 \rangle^
{1/2}=1.2 A^{1/3} \, \, fm$.  This approximation works very well for nuclei 
with $A \leq 100$ although it tends to overestimate the form factor for heavy 
nuclei.  All in all, we considered our approximations in nuclear physics
to be valid at the level of a few percent in nuclei with $A\leq 100$ 
(such as $Al$, $S$ and $Ti$), whereas for heavy nuclei (e.g $Pb$) they 
will give only the rough order of magnitude.

What is usually compared with experiment is not the absolute value of the
conversion rate but the ratio of this rate to the total muon capture rate, 
since this latter quantity is experimentally determined with very good 
precision \cite{suzuki}.  However for the rough orders of magnitude that 
we are interested in, we could also take the theoretical expression for 
the muon capture rate which has the virtue of eliminating in the ratio 
most of the nuclear dependence and can be applied to a wider range of 
elements where no accurate measurement is yet available.  The standard
semi-empirical expression for the nuclear muon capture rate is:
\cite{primakoff}:
\be
\Gamma_{cap.} = \frac{G_F^2 m_{\mu}^5 \alpha^3}{2 \pi^2}
\left( g_V^2 + 3 g_A^2 \right) 
Z_{eff}^4
\left(1 - \delta \frac{N}{2A}\right)
\label{capture}
\ee
where we can see the contribution of the Fermi and Gamow-Teller transition.
The last bracket, which is important for heavy nuclei, appears by virtue 
of the Pauli principle, namely that a proton which captures a muon must 
turn into a neutron and this it cannot do if the neutron state it would 
fill is already occupied by an existing neutron.  The empirical values for 
this parameter $\delta$ have been determined in \cite{schopper} and ranges
in value from 3.4 to 3.2 for medium to heavy nuclei.  With equations 
(\ref{muconversionrate}) and (\ref{capture}) we can write then the branching 
ratio as:
\be
B_{\mu-e} =   \frac{\Gamma_{conv.}}{\Gamma_{cap.}} =
\frac{1/(\zeta z)^4}{G_F^2} \frac{Q^2/Z}{(g_V^2 + 3 g_A^2)}
|F(q^2)|^2 \frac{1}{(1 - \delta \frac{N}{2A})}.
\label{ratio}
\ee
 
Finally, we turn to numerical results and their confrontation with experiment.
In Table 1, we list the branching ratios for coherent muon-electron conversion 
in several nuclei for which stringent limits are either already available
\cite{databook} or soon to be examined in planned experiments.  The 
theoretical conversion rates are calculated in DSM using the formulae 
(\ref{muconversionrate}), (\ref{m2}), and (\ref{formfactor}), with the 
{\it effective charges} $Z_{eff}$ taken from \cite{suzuki,shanker,chiang} 
and the nuclear form factor taken at the dominant kinematic point 
$q^2 \sim - m_{\mu}^2$.  The scale parameter $\zeta z$ is taken at 400 TeV 
in the middle of the range that we have estimated in the manner explained 
above, and the branching ratios are normalized to the experimentally 
measured values for $\Gamma_{cap.}$ \cite{suzuki} also listed.

\begin{table}
\begin{eqnarray*}
\begin{array}{||l|l|l|l|l||}
\hline \hline
Element        & Z_{eff} & \Gamma_{cap.}^{Exp.}(s^{-1}) & 
B_{\mu-e}^{Theor.} & B_{\mu-e}^{Exp. limit} \\
\hline
^{27} Al_{13}  & 11.48 & 0.66 \, \times \, 10^{6} & 2.8 \, \times \, 10^{-12} &
n. a.  \\
^{32} S_{16}   & 13.64 & 1.34 \, \times \, 10^{6} & 2.6 \, \times \, 10^{-12} &
7 \, \times \, 10^{-11}  \\
^{48} Ti_{22}  & 17.38 & 2.6 \, \times \, 10^{6}  & 4.6 \, \times \, 10^{-12} &
 4.3 \, \times \, 10^{-12}\\
^{207} Pb_{82} & 34.18 & 13.3 \, \times \, 10^{6} & 5.3 \, \times \, 10^{-12} &
4.6 \, \times \, 10^{-12} \\
\hline \hline
\end{array}
\end{eqnarray*}
\caption{Theoretical estimates for the ratio of the $\mu-e$ conversion rate 
to the $\mu$ capture rate compared with present experimental limits.  These
values are calculated with the scale parameter $\zeta z$ chosen at 400 TeV
at the middle of the estimated range, the same value as that used earlier
in \cite{FCNC} for calculating FCNC effects in meson mass differences and
rare meson decays .  For a discussion on the dependence of these results
on the scale $\zeta z$, see text.  The nuclear form factors were estimated 
using the dipole formula (\ref{formfactor}) except for $Pb$ for which a 
more detailed and realistic model was used \cite{pbformfactor}.  The more
accurate formula (\ref{chargeagain}) was used to estimate $Q$.}
\end{table}

In Table 1, one notes first that all the predicted branching ratios are 
roughly within the limits set by present experiment.  This we find very 
gratifying because, all the parameters in the scheme having already been 
fixed by earlier work in other physical contexts, there is no freedom 
left in the calculation.  A priori, no relation need exist linking to 
muon-electron conversion such widely different phenomena as quark masses, 
neutrino oscillations, meson decays, and cosmic ray air showers, from which 
our parameters were determined.  They are brought together in our case 
only through the DSM scheme.  If the scheme is altogether false, there 
would be no reason why our predictions here for muon-electron conversion 
may not come out many orders of magnitude outside the experimental bounds, 
and yet the answers in Table 1 all turn out to fall neatly within.  It is 
thus tempting to ascribe this coincidence to the DSM scheme's basic 
consistency.  Secondly, one notices that the predicted values are all 
close to the experimental limit in all cases where this is known, and thus 
should very soon be accessible to detection in the next generation of 
experiments.  Particularly interesting in this context are the cases of 
$^{27} Al_{13}$ \cite{meco} and $^{48} Ti_{22}$ \cite{psi} for which plans 
to increase the experimental sensitivity by 3 to 4 orders of magnitude are 
already underway.  

It should be noted, however, that the predictions given in Table 1 for the 
muon-electron conversion rate are very sensitive to the chosen value of the 
scale parameter $\zeta z$, occurring as it does in (\ref{muconversionrate})
as the fourth power.  We recall that the crucial upper bound on this 
parameter was deduced only through a somewhat qualitative argument from
a suggested neutrino explanation for post-GZK air showers \cite{airshower1},
and is thus unsure.  Nevertheless, if experimental sensitivity is improved
by 4 orders of magnitude as planned and still no muon-electron conversion
in nuclei is seen, it would mean pushing up our estimate of the scale 
parameter $\zeta z$ by another order of magnitude which would make the 
neutrino explanation for post-GZK air showers much less attractive,
perhaps even untenable.  This would be disappointing, for this suggestion
is so far the only direct test one has for the assumption that generation
is indeed dual color and not some other 3-fold horizontal gauge symmetry
\cite{phenodsm}, although such an outcome would still not invalidate the 
basic tenets of the DSM scheme.  On the other hand, if muon-electron 
conversion is observed with roughly the predicted rates, then one can
imagine turning the argument around and use the consequent estimate for
the scale $\zeta z$ to make definite predictions for neutrino interactions
at ultra-high energy, to be tested with post-GZK air showers, at the planned 
Auger project, for example.  However, a more serious scenario to the DSM 
scheme proper than a failure to observe any muon-electron conversion at 
roughly the predicted rate in Table 1 would be if muon-electron conversion 
is observed but turns out to have very different relative rates for the 
different nuclei than those predicted, for this would cast doubt either 
on the way the nuclear physics is handled above, or else on the whole 
mixing pattern of the DSM scheme from which the quark CKM matrix and 
neutrino oscillation parameters were evaluated.

The phenomenon of muon-electron conversion can of course be considered 
also in contexts other than coherent reactions in nuclei.  Another 
scenario in which it has been studied experimentally in some detail is 
in muonium-antimuonium conversion.  Using similar arguments based on the 
DSM scheme as those detailed above, but tailored to the atomic environment 
in muonium, we have also made a study of the conversion rate there. The 
value of the effective coupling relevant for this system in the DSM scheme
is:
$$
G_{DSM} \sim 2.5 \,\, \times \, \, 10^{-7} \, \, G_F,
$$
which in turn implies a conversion probability for muonium-antimuonium 
conversion integrated over all times of the order:
$$
P_{M \bar{M}} \sim 10^{-18}.
$$
At present, the best experimental bounds are in the range of $10^{-10}$
\cite{willmann1}, i.e. some 8 orders of magnitude above the predicted rates, 
which are thus presumably inaccessible for some time with the present 
experimental setup \cite{willmann2}.  For this reason the details of 
this analysis will not be presented here.

In summary, we conclude that the DSM scheme, as at present parametrized
and conceived, is consistent with the existing, already quite stringent, 
experimental bounds on muon-electron conversion, whether in nuclei or in
muonium, and that the predicted rates in coherent nuclear reactions are 
such as to make it very likely for the phenomenon to be discovered in 
the new generation of experiments now being planned.

\vspace{.5cm}

\noindent{\large {\bf Acknowledgment}}\\

One of us (JB) wishes to thank J. Bernab\'eu and F. Botella for very 
interesting discussions, E Oset for comments on muon capture and to 
acknowledge support from the Spanish Government on contract no. AEN 97-1718, 
PB97-1261 and GV98-1-80.

\end{document}